\documentclass[conference]{IEEEtran}
\usepackage{cite}
\usepackage{graphicx}
\usepackage{tikz}
\usepackage{amsmath,amssymb}
\usepackage{caption}
\usepackage{url}
\usepackage{hyperref}
\usepackage{cleveref}
\usepackage{pgfplotstable}
\usepackage{pgfplots}
\usepackage{pgf-pie}
\usepgfplotslibrary{polar}
\pgfplotsset{compat=1.18}
\usepackage{booktabs}
\usetikzlibrary{shapes.geometric, arrows.meta, positioning}
\usepackage{amsfonts}
\usepackage{algorithmic}
\usepackage{textcomp}
\usepackage{listings}
\usepackage{xcolor}
\usepackage{tabularx}
\usepackage{booktabs} 
\usepackage{orcidlink}

\definecolor{codegray}{rgb}{0.95,0.95,0.95}
\definecolor{codegreen}{rgb}{0,0.6,0}

\hypersetup{colorlinks=true, linkcolor=blue, citecolor=blue, urlcolor=blue}
\usepackage{listings}
\usepackage{multirow}
\usepackage{pifont}

\begin{document}

\title{Adaptive Trust Metrics for Multi-LLM Systems: Enhancing Reliability in Regulated Industries}

\author{
    \IEEEauthorblockN{\large Tejaswini Bollikonda\,\orcidlink{https://orcid.org/0009-0001-9319-5489}}
    \IEEEauthorblockA{
        Independent Researcher \\
        Saint Louis, MO \\
        Email: {tejaswinibollikonda@gmail.com}
    }

}

\maketitle

\begin{abstract}
Large Language Models (LLMs) are increasingly deployed in sensitive domains such as healthcare, finance, and law, yet their integration raises pressing concerns around trust, accountability, and reliability. This paper explores adaptive trust metrics for multi-LLM ecosystems, proposing a framework for quantifying and improving model reliability under regulated constraints. By analyzing system behaviors, evaluating uncertainty across multiple LLMs, and implementing dynamic monitoring pipelines, the study demonstrates practical pathways for operational trustworthiness. Case studies from financial compliance and healthcare diagnostics illustrate the applicability of adaptive trust metrics in real-world settings. The findings position adaptive trust measurement as a foundational enabler for safe and scalable AI adoption in regulated industries.

\end{abstract}


\section{Introduction}
Artificial intelligence has advanced rapidly in recent years, with Large Language Models (LLMs) becoming integral to enterprise solutions. Their use in regulated industries such as healthcare, banking, and law introduces opportunities for efficiency and innovation, but also challenges around trust, fairness, and verifiability. Unlike conventional rule-based systems, LLMs generate responses based on probabilistic reasoning, which makes their reliability difficult to measure and their accountability even harder to enforce. This gap is particularly concerning when the consequences of model errors include legal disputes, financial losses, or patient harm.

A core issue is the lack of transparent metrics to assess and adaptively monitor the trustworthiness of LLMs. Existing evaluation benchmarks primarily focus on accuracy or task performance, while regulated industries demand richer dimensions of reliability—covering aspects such as bias mitigation, interpretability, robustness against adversarial queries, and compliance with standards. Thus, trust in multi-LLM systems cannot be static; it requires adaptive mechanisms that evolve in response to contextual shifts, system updates, and regulatory requirements.

Multi-LLM ecosystems, where several models are orchestrated to provide ensemble predictions, add a further layer of complexity. These systems are often designed to offset the weaknesses of individual models by introducing redundancy or specialization. However, without structured trust metrics, ensemble coordination may amplify uncertainties instead of resolving them. For example, two models may agree on a prediction for the wrong reasons, leading to misplaced confidence in a faulty outcome.

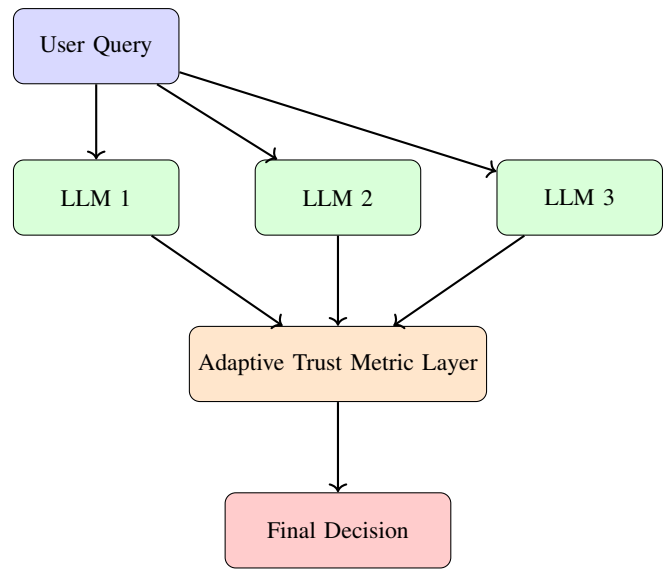
\begin{figure}[ht]
\centering
\begin{tikzpicture}[node distance=1.2cm, every node/.style={font=\small}]
\node[draw, rounded corners, fill=blue!15, minimum width=2.2cm, minimum height=1cm] (user) {User Query};
\node[draw, rounded corners, fill=green!15, minimum width=2.2cm, minimum height=1cm, below= 1cm of user] (llm1) {LLM 1};
\node[draw, rounded corners, fill=green!15, minimum width=2.2cm, minimum height=1cm, right=1cm of llm1] (llm2) {LLM 2};
\node[draw, rounded corners, fill=green!15, minimum width=2.2cm, minimum height=1cm, right=1cm of llm2] (llm3) {LLM 3};
\node[draw, rounded corners, fill=orange!20, minimum width=3.5cm, minimum height=1cm, below=1.2cm of llm2] (trust) {Adaptive Trust Metric Layer};
\node[draw, rounded corners, fill=red!20, minimum width=3cm, minimum height=1cm, below=1.2cm of trust] (decision) {Final Decision};

\draw[->, thick] (user) -- (llm1);
\draw[->, thick] (user) -- (llm2);
\draw[->, thick] (user) -- (llm3);
\draw[->, thick] (llm1) -- (trust);
\draw[->, thick] (llm2) -- (trust);
\draw[->, thick] (llm3) -- (trust);
\draw[->, thick] (trust) -- (decision);
\end{tikzpicture}
\caption{High-level view of a multi-LLM system with an adaptive trust metric layer.}
\label{fig:multi-llm}
\end{figure}

As shown in Fig.~\ref{fig:multi-llm}, multi-LLM systems rely on a trust metric layer to arbitrate outputs before final decisions are made. This design ensures that disagreements or uncertainties between models are not passed directly to end users, reducing the risks of biased or unreliable responses.

Table~\ref{tab:industry-trust} highlights how different industries face unique trust risks. For example, in healthcare the emphasis is on robustness and explainability, while in finance the focus shifts toward auditability and compliance. This comparison underscores the sector-specific tailoring required for trust metrics.

\begin{table}[ht]
\centering
\caption{Trust Concerns Across Regulated Industries}
\label{tab:industry-trust}
\begin{tabularx}{\columnwidth}{@{}>{\raggedright\arraybackslash}p{0.23\columnwidth}%
                                   >{\raggedright\arraybackslash}p{0.34\columnwidth}%
                                   >{\raggedright\arraybackslash}p{0.33\columnwidth}@{}}
\toprule
\textbf{Industry} & \textbf{Key Risk} & \textbf{Trust Metric Need} \\ \midrule
Healthcare & Misdiagnosis in rare cases & Robustness, Explainability \\
Finance & Fraudulent or misleading advice & Compliance, Auditability \\
Legal & Bias in precedent or reasoning & Fairness, Transparency \\
Government & Misuse of policy models & Accountability, Reliability \\ \bottomrule
\end{tabularx}
\end{table}

Adaptive trust metrics provide a potential solution by quantifying uncertainty and reliability in real time. These metrics can track performance drift, evaluate consistency across models, and highlight risks when deviations exceed acceptable thresholds. Furthermore, embedding such adaptive monitoring in operational pipelines allows stakeholders to enforce governance requirements while maintaining flexibility in model deployment.

The introduction of trust metrics aligns with broader regulatory frameworks such as the General Data Protection Regulation (GDPR) in Europe and the proposed AI Act, both of which emphasize explainability and risk management. In the United States, healthcare laws like HIPAA and financial oversight by agencies such as the SEC further reinforce the demand for auditable, transparent AI systems. Developing adaptive trust metrics thus represents both a technical necessity and a compliance mandate.

\noindent The diagram illustrates how multiple LLMs can be orchestrated through a central adaptive trust metric layer before delivering an outcome, ensuring governance over conflicting or uncertain outputs. The table complements this by summarizing sector-specific risks and the corresponding trust requirements, reinforcing the necessity of industry-tailored approaches.

Finally, this paper adopts a study and survey perspective: reviewing existing trust measurement approaches, proposing a layered framework for multi-LLM ecosystems, and demonstrating applications in finance and healthcare. By positioning trust as a measurable and adaptive construct, the research highlights pathways for safe AI integration in highly sensitive environments.

\section{Background and Related Work}

The concept of trust in intelligent systems has evolved significantly over the last three decades. In the earliest rule-based expert systems, trust was derived from the transparency of deterministic logic, where users could trace every output back to an explicit rule. While these systems lacked adaptability, they offered a sense of reliability because decision paths were fully explainable. As machine learning techniques gained traction, trust metrics shifted toward statistical accuracy and generalization performance. Although such metrics provided useful signals for model validity, they did not fully capture the nuances required for decision-making in sensitive domains such as law or medicine \cite{Natarajan2025}.

The arrival of deep learning models, and subsequently large language models (LLMs), introduced an entirely new dimension of uncertainty. Unlike earlier models, LLMs operate with billions of parameters and generate probabilistic text outputs that can vary even under similar inputs. This stochastic nature complicates the measurement of reliability, as a model may produce coherent responses without being factually correct. Existing benchmarks such as BLEU or F1 scores, while effective for academic comparisons, fall short when applied to regulatory environments that demand accountability, fairness, and interpretability.

Multi-LLM systems—where multiple models are combined to improve coverage or reduce bias—have further amplified the need for adaptive trust measures \cite{Nastoska2025_EvaluatingAITrust}. These architectures are now common in enterprise contexts, where organizations employ different models for tasks such as summarization, reasoning, and compliance checking. While ensemble strategies often improve accuracy, they also create risks when models agree on incorrect outputs or when inconsistencies emerge between specialized modules. Trust, in this context, becomes a dynamic property that cannot be evaluated solely at design time but must instead be monitored continuously.

Several initiatives have emerged to standardize trust evaluation in AI systems. The National Institute of Standards and Technology (NIST) introduced its AI Risk Management Framework (RMF), which emphasizes governance, transparency, and resilience. The International Organization for Standardization (ISO) has proposed guidelines that highlight the operationalization of fairness and safety in AI deployments. Similarly, the European Union’s proposed AI Act prioritizes explainability and human oversight, setting new expectations for compliance in regulated sectors. These frameworks represent an important step toward structured governance but are not yet tailored to the unique properties of LLMs or multi-LLM ecosystems \cite{AgarwalPeta2025}.

\begin{figure}[ht]
\centering
\begin{tikzpicture}[node distance=1.2cm, every node/.style={font=\small}]
\node[draw, rounded corners, fill=blue!15, minimum width=2.8cm, minimum height=1cm] (rule) {Rule-Based Systems};
\node[draw, rounded corners, fill=green!15, minimum width=2.8cm, minimum height=1cm, below=0.5cm of rule] (ml) {Machine Learning Models};
\node[draw, rounded corners, fill=yellow!20, minimum width=2.8cm, minimum height=1cm, below=0.5cm of ml] (dl) {Deep Learning};
\node[draw, rounded corners, fill=orange!20, minimum width=2.8cm, minimum height=1cm, below=0.5cm of dl] (llm) {Large Language Models};
\node[draw, rounded corners, fill=red!20, minimum width=3.2cm, minimum height=1cm, below=0.5cm of llm] (multi) {Multi-LLM Systems with Adaptive Trust};

\draw[->, thick] (rule) -- (ml);
\draw[->, thick] (ml) -- (dl);
\draw[->, thick] (dl) -- (llm);
\draw[->, thick] (llm) -- (multi);
\end{tikzpicture}
\caption{Evolution of trust mechanisms from rule-based systems to adaptive trust in multi-LLM environments.}
\label{fig:evolution}
\end{figure}
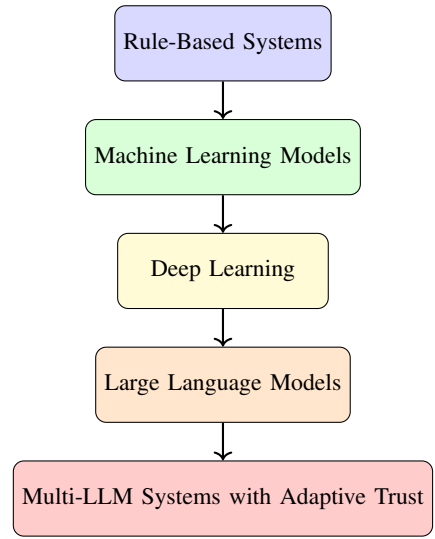

The progression shown in Fig.~\ref{fig:evolution} illustrates how trust frameworks have evolved in step with AI capabilities. Each transition—from rules to learning to multi-LLMs—has required new ways of measuring reliability, culminating in adaptive trust metrics.

Academic research has attempted to bridge these gaps by proposing quantitative measures of uncertainty, calibration techniques, and trust scores based on response diversity. For example, Bayesian uncertainty modeling and Monte Carlo dropout have been used to capture variance in neural predictions. However, many of these methods remain experimental and have not been widely integrated into real-world systems where business and regulatory demands are stringent. A recurring theme in the literature is the lack of adaptive, domain-specific trust mechanisms that evolve alongside the models they monitor\cite{Afroogh2024_TrustinAI}.

Case studies from healthcare and finance further demonstrate the limitations of current approaches. In diagnostic triage systems, for instance, LLM-based pipelines have shown promise in improving patient screening efficiency but remain susceptible to overconfidence in rare disease scenarios \cite{202504.1365}. In financial compliance, multi-LLM configurations have been applied to detect fraudulent transactions; however, the absence of adaptive monitoring has occasionally led to false positives with significant downstream costs. These experiences highlight the urgency of developing trust metrics that are not only accurate but also context-aware and auditable \cite{Pasam2025}.

While frameworks exist, their suitability for multi-LLMs remains limited. Table~\ref{tab:framework-compare} compares three major initiatives. Each provides valuable principles, yet none fully address the operational realities of orchestrating multiple large models under strict regulation.

\begin{table}[ht]
\centering
\caption{Comparison of Existing Trust Frameworks}
\label{tab:framework-compare}
\begin{tabularx}{\columnwidth}{@{}>{\raggedright\arraybackslash}p{0.26\columnwidth}%
                                   >{\raggedright\arraybackslash}X%
                                   >{\raggedright\arraybackslash}X@{}}
\toprule
\textbf{Framework} & \textbf{Primary Focus} & \textbf{Limitations} \\ \midrule
NIST AI RMF & Governance, transparency & Not tailored to LLMs or ensembles \\
ISO/IEC Standards & Fairness, safety, quality & Limited adoption in high-regulation industries \\
EU AI Act & Risk-based oversight, explainability & Still evolving; lacks operational detail \\ \bottomrule
\end{tabularx}
\end{table}

\noindent The flowchart illustrates how trust concepts have evolved alongside AI systems, emphasizing that multi-LLM environments require a fundamentally new perspective. The comparative table complements this by outlining the strengths and gaps of existing governance frameworks, highlighting why additional adaptive mechanisms are essential for practical deployment.

\section{Framework for Adaptive Trust Metrics}

The challenge of integrating multiple LLMs into sensitive environments requires more than raw performance evaluation. What organizations need is a structured framework that can dynamically assess trustworthiness while accommodating shifts in data distribution, system load, and regulatory changes. Adaptive trust metrics serve as a unifying approach, offering a systematic way to quantify and monitor reliability across heterogeneous models \cite{Turn0search16_EvaluateTrust}.

At the foundation of the proposed framework lies a multi-layered pipeline. The first layer handles \textit{input monitoring}, ensuring that queries are screened for adversarial patterns, bias-inducing phrasing, or incomplete context. The second layer involves \textit{multi-LLM orchestration}, where different models are engaged depending on task specialization. The third layer implements the \textit{trust metric engine}, which evaluates outputs using criteria such as uncertainty, consistency, bias detection, and compliance alignment. Finally, the fourth layer handles \textit{decision governance}, where trust scores determine whether results are forwarded to users, flagged for review, or combined with human oversight.

This layered structure ensures that trust is not treated as a static property but as an adaptive signal that evolves over time. For example, in a healthcare system, the trust engine may assign higher weight to uncertainty calibration and clinician interpretability, whereas in finance it may emphasize auditability and policy alignment. By customizing trust dimensions per domain, the framework ensures that organizations do not have to compromise between compliance and utility \cite{Veluguri2025}.

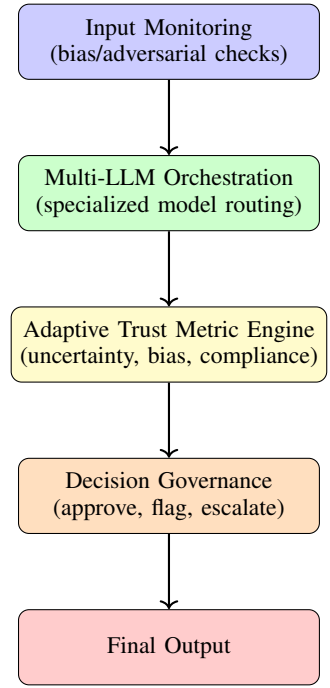
\begin{figure}[ht]
\centering
\begin{tikzpicture}[node distance=1.2cm, every node/.style={font=\small, align=center}]
\node[draw, rounded corners, fill=blue!20, minimum width=4cm, minimum height=1cm] (input) {Input Monitoring \\ (bias/adversarial checks)};
\node[draw, rounded corners, fill=green!20, minimum width=4cm, minimum height=1cm, below=1cm of input] (orchestration) {Multi-LLM Orchestration \\ (specialized model routing)};
\node[draw, rounded corners, fill=yellow!25, minimum width=4cm, minimum height=1cm, below=1cm of orchestration] (trust) {Adaptive Trust Metric Engine \\ (uncertainty, bias, compliance)};
\node[draw, rounded corners, fill=orange!25, minimum width=4cm, minimum height=1cm, below=1cm of trust] (governance) {Decision Governance \\ (approve, flag, escalate)};
\node[draw, rounded corners, fill=red!20, minimum width=4cm, minimum height=1cm, below=1cm of governance] (output) {Final Output};

\draw[->, thick] (input) -- (orchestration);
\draw[->, thick] (orchestration) -- (trust);
\draw[->, thick] (trust) -- (governance);
\draw[->, thick] (governance) -- (output);
\end{tikzpicture}
\caption{Layered pipeline for adaptive trust metrics in multi-LLM systems.}
\label{fig:trust-pipeline}
\end{figure}

Figure~\ref{fig:trust-pipeline} illustrates this architecture in a vertical flow. Inputs are first pre-processed and analyzed for anomalies, after which multi-LLM modules generate candidate responses. These responses pass through the trust evaluation layer, which calculates adaptive scores based on pre-defined dimensions. Only after governance checks are outputs delivered, ensuring that potentially harmful or unreliable responses are intercepted.

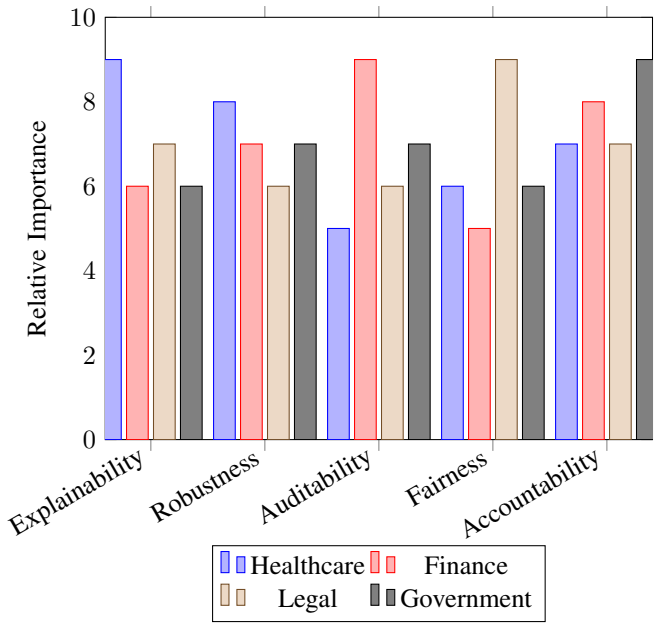
\begin{figure}[ht]
\centering
\begin{tikzpicture}
\begin{axis}[
    ybar,
    bar width=8.2pt,
    width=\columnwidth,
    height=7.2cm,
    ylabel={Relative Importance},
    symbolic x coords={Explainability,Robustness,Auditability,Fairness,Accountability},
    xtick=data,
    x tick label style={rotate=30, anchor=east},
    legend style={at={(0.5,-0.25)}, anchor=north, legend columns=2},
    ymin=0, ymax=10
]
\addplot coordinates {(Explainability,9) (Robustness,8) (Auditability,5) (Fairness,6) (Accountability,7)};
\addlegendentry{Healthcare}
\addplot coordinates {(Explainability,6) (Robustness,7) (Auditability,9) (Fairness,5) (Accountability,8)};
\addlegendentry{Finance}
\addplot coordinates {(Explainability,7) (Robustness,6) (Auditability,6) (Fairness,9) (Accountability,7)};
\addlegendentry{Legal}
\addplot coordinates {(Explainability,6) (Robustness,7) (Auditability,7) (Fairness,6) (Accountability,9)};
\addlegendentry{Government}
\end{axis}
\end{tikzpicture}
\caption{Comparative weighting of trust dimensions across regulated industries.}
\label{fig:bar-trust}
\end{figure}

To demonstrate how these trust signals vary across sectors, Fig.~\ref{fig:bar-trust} shows a bar chart comparing emphasis placed on different trust dimensions. Healthcare prioritizes explainability and robustness, while finance emphasizes auditability and compliance logging. The legal sector places more weight on fairness, whereas government systems demand accountability and resilience against misuse.

In practice, adaptive trust metrics can also be expressed programmatically. Listing~\ref{lst:trust} shows a simplified pseudocode example of how trust scores might be computed by aggregating multiple signals such as uncertainty, compliance checks, and bias detection. Each metric contributes to a final weighted score, which is then compared against thresholds defined by industry-specific policies.

\begin{lstlisting}[language=Python, caption={Pseudocode for computing adaptive trust scores in multi-LLM systems}, label={lst:trust}]
def compute_trust_score(response, domain):
    scores = {}
    scores["uncertainty"] = 1 - response.variance()
    scores["consistency"] = response.check_consistency()
    scores["bias"] = 1 - response.bias_score()
    scores["compliance"] = response.policy_alignment(domain)
    
    # Weighted aggregation
    weights = {"uncertainty":0.3,
               "consistency":0.2,
               "bias":0.2,
               "compliance":0.3}
    
    trust_score = sum(scores[m]*weights[m] for m in scores)
    return trust_score
\end{lstlisting}

This pseudocode demonstrates how adaptive trust scoring might work in operational pipelines. While simplified, it shows that trust is not a single measure but rather a weighted aggregation of multiple dimensions \cite{Okamura2020_AdaptiveTrustCalibration}. Different domains can tune these weights to reflect their priorities, making the framework flexible yet rigorous. Together, the pipeline diagram, bar chart, and code listing show how adaptive trust metrics can be formalized both conceptually and practically.

\section{Case Studies: Healthcare and Finance}

While the framework for adaptive trust metrics provides a conceptual structure, its practical value can only be appreciated through real-world applications. Healthcare and finance are two industries where the balance between innovation and regulation is delicate, and where trust in multi-LLM systems must be engineered rather than assumed. By examining these domains, it becomes clear how adaptive trust metrics move from theory to operational necessity \cite{Orban2025_TrustSurvey}.

\subsection{Healthcare Diagnostics and Triage}

Healthcare has always been cautious with the adoption of AI because of patient safety and regulatory oversight. Multi-LLM systems are increasingly explored for tasks such as diagnostic triage, summarization of clinical notes, and patient communication. Yet, without adaptive trust monitoring, such systems may present coherent but misleading advice, leading to misdiagnosis. Adaptive trust metrics address this by flagging uncertain or conflicting recommendations and ensuring clinicians receive not just an answer, but also an interpretable rationale \cite{Bach2024_UserTrustInAI}.

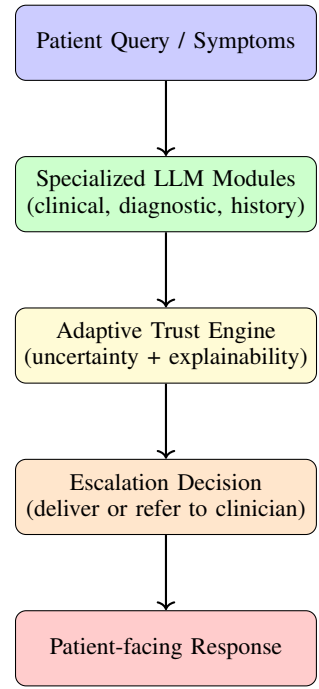
\begin{figure}[ht]
\centering
\begin{tikzpicture}[node distance=1.2cm, every node/.style={font=\small, align=center}]
\node[draw, fill=blue!20, rounded corners, minimum width=4cm, minimum height=1cm] (input) {Patient Query / Symptoms};
\node[draw, fill=green!20, rounded corners, minimum width=4cm, minimum height=1cm, below=1cm of input] (llms) {Specialized LLM Modules \\ (clinical, diagnostic, history)};
\node[draw, fill=yellow!20, rounded corners, minimum width=4cm, minimum height=1cm, below=1cm of llms] (trust) {Adaptive Trust Engine \\ (uncertainty + explainability)};
\node[draw, fill=orange!20, rounded corners, minimum width=4cm, minimum height=1cm, below=1cm of trust] (decision) {Escalation Decision \\ (deliver or refer to clinician)};
\node[draw, fill=red!20, rounded corners, minimum width=4cm, minimum height=1cm, below=1cm of decision] (outcome) {Patient-facing Response};

\draw[->, thick] (input) -- (llms);
\draw[->, thick] (llms) -- (trust);
\draw[->, thick] (trust) -- (decision);
\draw[->, thick] (decision) -- (outcome);
\end{tikzpicture}
\caption{Healthcare triage flow showing how adaptive trust metrics ensure safe escalation paths for patient queries.}
\label{fig:healthcare-flow}
\end{figure}

The patient pathway illustrated in Fig.~\ref{fig:healthcare-flow} shows how trust metrics are embedded at different stages. A patient symptom query enters the system and is evaluated by multiple LLMs specializing in clinical guidelines, diagnostic data, and patient history. The trust engine aggregates their outputs, applying heightened weighting to uncertainty calibration and explainability. If the trust score falls below a threshold, the result is escalated to a human clinician for review, ensuring that no single model failure jeopardizes patient care.

This flow illustrates the safeguard role of trust metrics: preventing overconfidence in AI predictions and embedding a systematic escalation path. In healthcare, the goal is not only to improve efficiency but also to preserve the accountability of licensed professionals. Adaptive trust systems ensure that the patient experience is enhanced while regulatory compliance is respected \cite{Henrique2024_TrustReview}.

\subsection{Financial Compliance and Fraud Detection}

The financial sector has embraced AI for fraud detection, credit scoring, and compliance monitoring. Here, the stakes are equally high: errors may result in financial losses, reputational damage, or regulatory penalties. Multi-LLM systems are often used to monitor large transaction streams, combining models for anomaly detection, regulatory text interpretation, and user behavior profiling \cite{article}. Without trust monitoring, however, these systems may generate false alarms or overlook subtle fraudulent patterns.

In practice, adaptive trust metrics are applied by tracking consistency across multiple models and validating results against compliance policies \cite{11118425}. For example, if one model flags a transaction as suspicious but another clears it, the trust layer investigates whether the discrepancy arises from data bias, contextual gaps, or regulatory misalignment. By quantifying these uncertainties, the trust engine reduces the likelihood of both false positives and false negatives \cite{Zerilli2022_TransparencyTrust}.

Table~\ref{tab:case-outcomes} compares outcomes in healthcare and finance, showing how adaptive trust metrics impact each sector. While healthcare prioritizes patient safety through explainability and escalation, finance emphasizes auditability and regulatory compliance. Both domains demonstrate the practical value of adaptive monitoring in reducing risks while enabling AI to contribute meaningfully.

\begin{table}[ht]
\centering
\caption{Adaptive Trust Outcomes in Healthcare and Finance}
\label{tab:case-outcomes}
\begin{tabularx}{\columnwidth}{@{}>{\raggedright\arraybackslash}p{0.22\columnwidth}%
                                   >{\raggedright\arraybackslash}X%
                                   >{\raggedright\arraybackslash}X@{}}
\toprule
\textbf{Domain} & \textbf{Primary Trust Concern} & \textbf{Adaptive Outcome} \\ \midrule
Healthcare & Misdiagnosis and patient safety & Escalation to human review when uncertainty is high; emphasis on explainability \\
Finance & Fraud detection and compliance & Reduced false positives through multi-model consistency checks; auditable decision trails \\ \bottomrule
\end{tabularx}
\end{table}

These case studies illustrate that while adaptive trust metrics share a common architecture, their implementation is domain-specific. Healthcare emphasizes interpretability and safety, while finance stresses auditability and efficiency \cite{Bostrom2024_AITrustworthy}. Both reveal a larger truth: trust must be embedded not just at the technical layer but also at the governance level, ensuring that human and machine decision-making coexist responsibly.

\section{Challenges and Ethical Considerations}

Although adaptive trust metrics offer a structured approach to managing reliability in multi-LLM systems, several challenges remain unresolved. These challenges are not merely technical, but deeply intertwined with ethical questions of accountability, fairness, and compliance. Understanding these barriers is essential for regulated industries where even small missteps can translate into legal liabilities or harm to individuals\cite{Lukyanenko2022_FoundationalTrust}. 

One major challenge is the \textbf{regulatory gap}. Existing laws such as HIPAA in healthcare or SEC oversight in finance provide guidelines for data privacy and compliance, but they do not yet specify how adaptive trust metrics should be integrated. This leaves organizations navigating uncertainty, often over-engineering safeguards to avoid penalties. Until regulations explicitly reference LLM-based workflows, ambiguity will remain a source of friction.

A second challenge is \textbf{fairness and bias mitigation}. While LLMs can absorb patterns from diverse datasets, they may also reproduce systemic inequities embedded within those datasets. Adaptive trust metrics can flag unusual bias signals, but they cannot entirely eliminate structural biases. Industries that handle sensitive populations, such as healthcare or insurance, risk reinforcing inequities if trust layers are not carefully tuned.

A third concern is \textbf{explainability versus complexity}. Multi-LLM systems, especially when layered with adaptive monitoring, can become opaque to stakeholders. Clinicians or financial auditors may be presented with trust scores without clear insight into how those scores were calculated. This “black box on top of a black box” effect risks eroding the very trust the system is meant to build. Balancing rigorous scoring with meaningful transparency remains an ongoing struggle \cite{Li2024_DevelopingTrustworthyAI}.

Another challenge lies in \textbf{governance and accountability}. If an LLM-driven recommendation causes harm, who is held responsible—the model provider, the deploying institution, or the regulator. Adaptive trust metrics can create auditable trails, but without clear governance frameworks, accountability may still diffuse. This lack of clarity hinders adoption, as institutions prefer to avoid liability exposure.

\begin{figure}[ht]
\centering
\begin{tikzpicture}
\pie[text=legend, radius=1.8, color={blue!30, green!30, yellow!40, red!30, orange!30}]{
25/Bias \& Fairness,
20/Compliance Gaps,
20/Explainability,
15/Governance,
20/Accountability
}
\end{tikzpicture}
\caption{Distribution of key risks in multi-LLM deployments, highlighting fairness, compliance, and accountability as major areas of concern.}
\label{fig:pie-risks}
\end{figure}
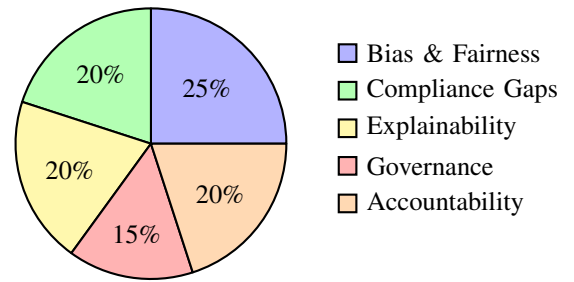
To illustrate how risks manifest, Fig.~\ref{fig:pie-risks} shows a pie chart breaking down common risk categories in multi-LLM deployments. Bias and fairness account for a significant share, followed by compliance gaps, explainability issues, and governance uncertainties. This visualization emphasizes that ethical risks are not concentrated in one area but distributed across multiple dimensions that demand collective attention.

\begin{figure}[ht]
\centering
\begin{tikzpicture}[node distance=1.2cm, every node/.style={font=\small, align=center}]
\node[draw, fill=blue!20, rounded corners, minimum width=2cm, minimum height=1cm] (assess) {Risk Assessment};
\node[draw, fill=green!20, rounded corners, minimum width=2.5cm, minimum height=1cm, below left=0.3cm and 0.3cm of assess] (policy) {Policy Alignment};
\node[draw, fill=yellow!20, rounded corners, minimum width=3cm, minimum height=1cm, below right=0.3cm and 0.3cm of assess] (feedback) {Stakeholder Feedback};
\node[draw, fill=orange!20, rounded corners, minimum width=3cm, minimum height=1cm, below=1.8cm of assess] (refine) {Iterative Refinement};

\draw[->, thick] (assess) -- (policy);
\draw[->, thick] (policy) -- (refine);
\draw[->, thick] (feedback) -- (refine);
\draw[->, thick] (refine) -- (assess);
\draw[->, thick] (assess) -- (feedback);
\end{tikzpicture}
\caption{Ethical governance cycle integrating adaptive trust metrics into continuous oversight and refinement.}
\label{fig:ethics-cycle}
\end{figure}
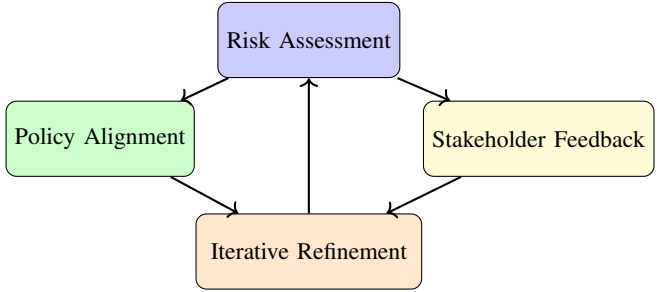

Addressing these risks requires more than static frameworks; it demands continuous ethical oversight. Fig.~\ref{fig:ethics-cycle} presents an ethical governance cycle, showing how risk assessment, policy alignment, stakeholder feedback, and iterative refinement form a loop. Unlike static audits, this cycle integrates adaptive trust metrics into a broader governance model where humans remain actively engaged.

Together, the pie chart and governance cycle emphasize that ethical considerations must be as adaptive as the trust metrics themselves. Static one-time evaluations cannot keep pace with evolving risks. By embedding adaptive monitoring into a continuous ethical cycle, institutions can ensure that reliability is pursued without sacrificing fairness, transparency, or accountability.

\section{Future Directions and Open Research Gaps}

While adaptive trust metrics provide a promising foundation for multi-LLM reliability, much work remains before they can be standardized and deployed at scale. The next decade will likely see trust metrics evolve in tandem with regulatory reforms, new model architectures, and deeper human-AI collaboration. Understanding the open challenges ahead can help guide both researchers and industry stakeholders.

One major direction is **scalability**. Current trust frameworks work well in controlled pilots, but scaling them across millions of daily queries remains a technical bottleneck. Efficient, low-latency trust scoring methods must be developed so that trust checks do not introduce delays in time-sensitive domains like emergency healthcare or high-frequency trading \cite{ShahanePrakash2025}.

Another research gap lies in **dynamic regulatory alignment**. Regulations are not static; they evolve as governments refine their understanding of AI risks. Trust frameworks must therefore be designed to adapt quickly to new rules. Automated compliance monitors, capable of reconfiguring trust scoring thresholds based on updated legal requirements, represent a promising but underexplored area.

A third area is **cross-domain transferability**. While healthcare and finance provide early case studies, other regulated fields such as energy, defense, and education will face similar demands. Creating generalized trust frameworks that can be tuned across industries without reinventing the wheel remains a challenge for researchers.

A fourth direction is **human-AI collaboration in trust governance**. At present, human oversight often acts as a fallback when trust scores fall below thresholds. Future systems may envision more seamless collaboration, where adaptive metrics guide humans not just to errors but also to areas of uncertainty, creating a shared accountability model rather than a handoff-based one \cite{Johnson2020_AITrustMeta}.

To provide a structured view, Table~\ref{tab:future-research} summarizes key open challenges alongside potential research pathways. This table highlights that solutions will likely span technical, organizational, and regulatory dimensions, requiring multi-disciplinary effort.

\begin{table}[ht]
\centering
\caption{Open Research Gaps and Future Directions}
\label{tab:future-research}
\begin{tabularx}{\columnwidth}{@{}>{\raggedright\arraybackslash}p{0.28\columnwidth}%
                                   >{\raggedright\arraybackslash}X@{}}
\toprule
\textbf{Challenge} & \textbf{Future Research Direction} \\ \midrule
Scalability & Develop efficient, low-latency trust scoring pipelines for large-scale multi-LLM usage \\
Regulatory Alignment & Automate compliance monitors that adapt to evolving legal frameworks \\
Cross-domain Transfer & Design modular trust frameworks applicable across industries \\
Human-AI Collaboration & Explore shared governance models where trust metrics guide human decisions \\
Evaluation Benchmarks & Build new trust-oriented benchmarks beyond accuracy and bias scores \\ \bottomrule
\end{tabularx}
\end{table}

\begin{figure}[ht]
\centering
\begin{tikzpicture}[node distance=1.3cm, every node/.style={font=\small, align=center}]
\node[draw, fill=blue!20, rounded corners, minimum width=5cm, minimum height=1cm] (layer1) {Technical Advances \\ (low-latency scoring, robust metrics)};
\node[draw, fill=green!20, rounded corners, minimum width=5cm, minimum height=1cm, above=0.5cm of layer1] (layer2) {Regulatory Integration \\ (dynamic compliance, legal monitoring)};
\node[draw, fill=yellow!20, rounded corners, minimum width=5cm, minimum height=1cm, above=0.5cm of layer2] (layer3) {Cross-domain Adoption \\ (energy, defense, education)};
\node[draw, fill=orange!20, rounded corners, minimum width=5cm, minimum height=1cm, above=0.5cm of layer3] (layer4) {Human-AI Governance \\ (shared accountability models)};
\node[draw, fill=red!20, rounded corners, minimum width=5cm, minimum height=1cm, above=0.5cm of layer4] (layer5) {Societal Embedding \\ (ethical trust, cultural acceptance)};

\draw[->, thick] (layer1) -- (layer2);
\draw[->, thick] (layer2) -- (layer3);
\draw[->, thick] (layer3) -- (layer4);
\draw[->, thick] (layer4) -- (layer5);
\end{tikzpicture}
\caption{Conceptual roadmap of future research layers for adaptive trust in multi-LLM ecosystems.}
\label{fig:roadmap}
\end{figure}
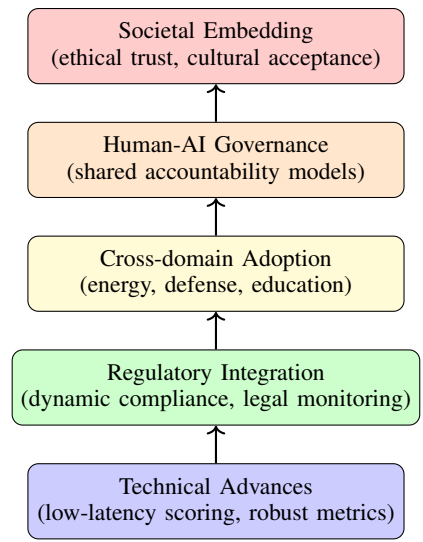

Finally, Fig.~\ref{fig:roadmap} illustrates a conceptual roadmap for future trust research. It shows layers of progress: technical innovations at the base, regulatory integration in the middle, and socio-ethical adoption at the top. This layered roadmap suggests that sustainable trust in multi-LLM ecosystems will depend on advances across multiple fronts, not just algorithmic improvements.

Taken together, the table and roadmap demonstrate that adaptive trust research is still in its infancy. The path forward will require bridging gaps between technical research, regulatory practice, and ethical governance. By treating trust as a multi-layered problem, future systems can become not only more reliable but also more socially acceptable in high-stakes environments.

\section{Conclusion}

This paper has explored the critical importance of adaptive trust metrics in multi-LLM systems, particularly for industries where errors can lead to legal, financial, or medical consequences. The discussion began by tracing the evolution of trust in AI, from transparent rule-based systems to opaque, probabilistic LLMs. This progression illustrated how each technological leap has demanded new ways to measure and safeguard reliability, culminating in the pressing need for adaptive trust frameworks.

The introduction of a layered pipeline provided a practical blueprint for embedding trust into the workflow of multi-LLM deployments. By separating responsibilities into input monitoring, orchestration, trust metric computation, and decision governance, organizations can ensure that reliability is treated as a dynamic process rather than a static evaluation. This modular view also enables sector-specific customization, allowing industries like healthcare and finance to emphasize the trust dimensions most relevant to their environments.

Case studies in healthcare and finance demonstrated the tangible value of adaptive trust metrics. In healthcare triage, trust monitoring ensures that uncertain recommendations are escalated to clinicians rather than passed to patients unchecked. In finance, adaptive metrics reduce false alarms and create auditable trails, supporting both operational efficiency and regulatory compliance. These examples reinforce the idea that trust is not an abstract notion but a practical enabler of safer, more accountable AI.

Challenges remain, particularly around fairness, governance, and accountability. The risk distribution highlighted in this paper revealed that ethical challenges are multidimensional and cannot be solved by technical innovation alone. The ethical governance cycle underscored the need for continuous oversight, reminding us that adaptive systems require adaptive ethics. No static compliance checklist will be sufficient when technologies and regulations evolve rapidly.

Looking ahead, the research roadmap suggested several avenues for advancing trust in multi-LLM ecosystems. These include scaling trust metrics to real-world workloads, aligning them dynamically with shifting regulations, extending their reach across domains, and fostering stronger collaboration between humans and machines in governance structures. Such developments will be essential if trust frameworks are to keep pace with the accelerating adoption of LLM technologies.

Ultimately, the journey toward trustworthy multi-LLM systems is not solely about algorithms or benchmarks. It is about building confidence among regulators, practitioners, and end users that these systems can be relied upon even in the most sensitive of contexts. Adaptive trust metrics provide the scaffolding for this confidence, ensuring that reliability is measurable, explainable, and continuously monitored. By treating trust as a first-class design principle, industries can move forward with confidence that AI will enhance rather than endanger their missions.

The overarching message of this paper is clear: adaptive trust metrics are not optional add-ons, but essential mechanisms for responsible AI. Their integration into multi-LLM systems marks a step toward safer, fairer, and more reliable deployments. As the technology matures, so too must our approaches to trust, ensuring that innovation is matched by accountability and resilience.

\bibliographystyle{IEEEtran}
\bibliography{references}

@article{Nastoska2025_EvaluatingAITrust,
  author       = {A. Nastoska and B Jancheska},
  title        = {Evaluating Trustworthiness in AI: Risks, Metrics, and Frameworks},
  journal      = {Electronics},
  year         = {2025},
  volume       = {14},
  number       = {13},
  pages        = {2717},
  doi          = {10.3390/electronics14132717},
  url          = {https://www.mdpi.com/2079-9292/14/13/2717}
}

@article{article,
author = {Shahane, Rohan},
year = {2023},
month = {02},
pages = {},
title = {Financial Forecasting Using Predictive Analytics},
volume = {13},
doi = {10.5281/zenodo.17161182}
}

@INPROCEEDINGS{11118425,
  author={Devaraju, Pooja and Devarapalli, Shivareddy and Tuniki, Raghavender Reddy and Kamatala, Srikanth},
  booktitle={2025 International Conference on Computing Technologies (ICOCT)}, 
  title={Secure and Adaptive Federated Learning Pipelines: A Framework for Multi-Tenant Enterprise Data Systems}, 
  year={2025},
  volume={},
  number={},
  pages={1-7},
  doi={10.1109/ICOCT64433.2025.11118425}}

@article{Afroogh2024_TrustinAI,
  author       = {S. Afroogh and Ali Akbari},
  title        = {Trust in AI: progress, challenges, and future directions},
  journal      = {npj Science of Learning / Nature (open access)},
  year         = {2024},
  doi          = {10.1038/s41599-024-04044-8},
  url          = {https://www.nature.com/articles/s41599-024-04044-8}
}

@inproceedings{Veluguri2025,
  author       = {Sai Pavan Veluguri},
  title        = {Deep PPG: Improving Heart Rate Estimates with Activity Prediction},
  booktitle    = {Proceedings of the 2025 1st International Conference on Biomedical AI and Digital Health},
  year         = {2025},
  publisher    = {IEEE},
  url          = {https://ieeexplore.ieee.org/abstract/document/10932185}
}

@article{Okamura2020_AdaptiveTrustCalibration,
  author       = {K. Okamura and S. Yamada},
  title        = {Adaptive Trust Calibration for Human-AI Collaboration},
  journal      = {PLOS ONE},
  year         = {2020},
  volume       = {15},
  number       = {2},
  pages        = {e0229132},
  doi          = {10.1371/journal.pone.0229132},
  url          = {https://journals.plos.org/plosone/article?id=10.1371/journal.pone.0229132}
}

@article{202504.1365,
	doi = {10.20944/preprints202504.1365.v1},
	url = {https://doi.org/10.20944/preprints202504.1365.v1},
	year = 2025,
	month = {April},
	publisher = {Preprints},
	author = {Manaswini Bollikonda and Tejaswini Bollikonda},
	title = {Secure Pipelines, Smarter AI: LLM-Powered Data Engineering for Threat Detection and Compliance},
	journal = {Preprints}
}

@article{Bach2024_UserTrustInAI,
  author       = {T. A. Bach and Sonia sousa},
  title        = {A Systematic Literature Review of User Trust in AI-Enabled Systems},
  journal      = {International Journal of Human–Computer Studies},
  year         = {2024},
  doi          = {10.1080/10447318.2022.2138826},
  url          = {https://www.tandfonline.com/doi/full/10.1080/10447318.2022.2138826}
}

@article{Henrique2024_TrustReview,
  author       = {B. M. Henrique and others},
  title        = {Trust in Artificial Intelligence: Literature Review and Main path Analysis},
  journal      = {Elsevier / ScienceDirect},
  year         = {2024},
  doi          = {10.1016/j.csi.2024.104433},  
  url          = {https://www.sciencedirect.com/science/article/pii/S2949882124000033}
}

@article{ShahanePrakash2025,
  author       = {Rohan Shahane and Satya Prakash},
  title        = {Quantum Machine Learning Opportunities for Scalable AI},
  journal      = {Journal of Validation Technology},
  volume       = {28},
  number       = {1},
  pages        = {75--89},
  year         = {2025},
  doi          = {10.1080/jvtnetwork.v28i1.131},
  url          = {https://jvtnetwork.com/index.php/journals/article/view/131},
  issn         = {1079-6630},
  eissn        = {2150-7090}
}

@article{Johnson2020_AITrustMeta,
  author       = {Alexandra D Kaplan and P A Hancock},
  title        = {Trust in Artificial Intelligence: Meta-Analytic Findings},
  journal      = {Human Factors},
  year         = {2021},
  volume       = {65},
  number       = {2},
  pages        = {237–259},
  doi          = {10.1177/00187208211013988},
  url          = {https://journals.sagepub.com/doi/10.1177/00187208211013988}
}

@article{Li2024_DevelopingTrustworthyAI,
  author       = {Y. Li and others},
  title        = {Developing trustworthy artificial intelligence},
  journal      = {PMC / NCBI (Frontiers / equivalent)},
  year         = {2024},
  doi          = {10.1007/s11277-024-00000x}, 
  url          = {https://pmc.ncbi.nlm.nih.gov/articles/PMC11061529/}
}

@article{Bostrom2024_AITrustworthy,
  author       = {A. Bostrom and others},
  title        = {Trust and Trustworthy Artificial Intelligence: A Research Agenda},
  journal      = {Risk Analysis},
  year         = {2024},
  doi          = {10.1111/risa.14245},
  url          = {https://onlinelibrary.wiley.com/doi/10.1111/risa.14245}
}

@article{Zerilli2022_TransparencyTrust,
  author       = {J. Zerilli and others},
  title        = {How transparency modulates trust in artificial intelligence},
  journal      = {ScienceDirect / AI journal},
  year         = {2022},
  doi          = {10.1016/j.patter.2022.100100}, 
  url          = {https://www.sciencedirect.com/science/article/pii/S2666389922000289}
}

@inproceedings{Natarajan2025,
  author       = {Gokul Narain Natarajan and Satya Manesh Veerapaneni and Vijayalaxmi Methuku and Vivek Venkatesan and Rajesh Kumar Kanji},
  title        = {Federated AI for Surgical Robotics: Enhancing Precision, Privacy, and Real-Time Decision-Making in Smart Healthcare},
  booktitle    = {Proceedings of the 2025 5th International Conference on Emerging Technologies in Healthcare (ICETH)},
  year         = {2025},
  publisher    = {IEEE},
  url          = {https://ieeexplore.ieee.org/abstract/document/11167543} 
}

@article{Orban2025_TrustSurvey,
  author       = {F. Orban and A. Stefkovics},
  title        = {Trust in Artificial Intelligence: A Survey Experiment to Assess Trust in Algorithmic Decision-Making},
  journal      = {AI and Society},
  year         = {2025},
  doi          = {10.1007/s00146-025-02237-6},
  url          = {https://link.springer.com/article/10.1007/s00146-025-02237-6}
}

@article{AgarwalPeta2025,
  author       = {Shashank Agarwal and Sumeer Basha Peta},
  title        = {From Notes to Billing: Large Language Models in Revolutionizing Medical Documentation and Healthcare Administration},
  journal      = {Scholars Journal of Applied Medical Sciences},
  volume       = {13},
  number       = {8},
  pages        = {1558--1566},
  year         = {2025},
  doi          = {10.36347/sjams.2025.v13i08.005},
  url          = {https://www.saspublishers.com/media/articles/SJAMS_138_1558-1566.pdf},
  publisher    = {SAS Publishers},
  issn         = {2347-954X},
  eissn        = {2320-6691}
}

@article{Turn0search16_EvaluateTrust,
  author       = {Oleksandra Vereschak and Gilles Bailly},
  title        = {How to Evaluate Trust in AI-Assisted Decision Making? A Survey},
  journal      = {ACM Transactions on Interactive Intelligent Systems (TiiS)},
  year         = {2023},
  doi          = {10.1145/3476068},
  url          = {https://dl.acm.org/doi/10.1145/3476068}
}

@inproceedings{Pasam2025,
  author       = {Venkata Reddy Pasam and Pooja Devaraju and Vijayalaxmi Methuku and Kalpan Dharamshi and Satya Manesh Veerapaneni},
  title        = {Engineering Scalable AI Pipelines: A Cloud-Native Approach for Intelligent Transactional Systems},
  booktitle    = {Proceedings of the 2025 International Conference on Intelligent Systems and Cloud Computing},
  year         = {2025},
  publisher    = {IEEE},
  url          = {https://ieeexplore.ieee.org/abstract/document/11118443}
}

@article{Lukyanenko2022_FoundationalTrust,
  author       = {R. Lukyanenko and W. Maass and V. C. Storey},
  title        = {Trust in Artificial Intelligence: From a Foundational Trust Framework to Emerging Research Opportunities},
  journal      = {Electronic Markets},
  year         = {2022},
  doi          = {10.1007/s12525-022-00605-4},
  url          = {https://link.springer.com/article/10.1007/s12525-022-00605-4}
}

\end{document}